\documentclass[runningheads]{llncs}
\usepackage[margin=2.5cm]{geometry}
\usepackage{color}
\usepackage{graphicx}
\usepackage{amsfonts}
\usepackage{latexsym,amssymb,amsmath}

\usepackage[ruled]{algorithm}
\usepackage[noend]{algpseudocode}
\algdef{SE}[DOWHILE]{Do}{doWhile}{\algorithmicdo}[1]{\algorithmicwhile\ #1}
\usepackage{comment}
\usepackage{epstopdf}
\usepackage{soul}
\usepackage[hidelinks]{hyperref}

\usepackage{lipsum} 
\usepackage{xargs} 

\usepackage{multirow}
\usepackage[pdftex,dvipsnames,table,xcdraw]{xcolor}

\usepackage[appendix=inline]{apxproof}

\authorrunning{J. Coleman et al.}

\newtheoremrep{lemma}{Lemma}[section]{\bfseries}{\itshape}
\newtheoremrep{theorem}{Theorem}[section]{\bfseries}{\itshape}

\begin{document}

\title{% 
    The Pony Express Communication Problem% 
    \thanks{This is the full version of a paper to appear at IWOCA 2021}
}

\author{
Jared Coleman\inst{1}
\and
Evangelos Kranakis\inst{3}\inst{5}
\and
Danny Krizanc\inst{4}
\and
Oscar Morales Ponce\inst{2}
}

\institute{
Department of Computer Science,
University of Southern California, California, USA \\
\email{jaredcol@usc.edu}
\and
School of Computer Science, 
Carleton University, Ottawa, Ontario, Canada \\
\email{kranakis@scs.carleton.ca}
\and
Department of Mathematics \& Computer Science, 
Wesleyan University, Middletown CT, USA \\
\email{dkrizanc@wesleyan.edu}
\and
Department of Computer Engineering and Computer Science, 
California State University, Long Beach, USA \\
\email{Oscar.MoralesPonce@csulb.edu}
\and
Research supported in part by NSERC Discovery grant.
}

\setcounter{tocdepth}{10}
\maketitle
\begin{abstract}
We introduce a new problem which we call the Pony Express problem. $n$ robots with differing speeds
are situated over some domain.
A message is placed at some commonly known point. 
Robots can acquire the message either by visiting its initial position, or by encountering another robot that has already acquired it.
The robots must collaborate to deliver the message to a given destination%
\footnote{We restrict our attention to message transmission rather than package delivery, which differs from message transmission in that packages cannot be replicated.}.%
The objective is to deliver the message in minimum time.
In this paper we study the Pony Express problem on the line where $n$ robots are arbitrarily deployed along a finite segment. 
%The robots have different speeds and can move in both directions on the line. 
We are interested in both offline centralized and online distributed algorithms.  
In the online case, we assume the robots have limited knowledge of the initial configuration.
In particular, the robots do not know the initial positions and speeds of the other robots nor even their own position and speed. They do, however, know the direction on the line in which to find the message and have the ability to compare speeds when they meet.

First, we study the Pony Express problem where the message is initially placed at one endpoint (labeled 0) of a segment
and must be delivered to the other endpoint (labeled 1).
We provide an $O(n \log n)$ running time offline algorithm as well as an optimal (competitive ratio 1) online algorithm. 
Then we study the Half-Broadcast problem where the message is at the center (at 0) and must be delivered to either one of the endpoints of the
segment $[-1,+1]$. 
We provide an offline algorithm running in $O(n^2 \log n)$ time
and we provide an online algorithm that attains a competitive ratio of $\frac 32$ which we show is the best possible. 
Finally, we study the Broadcast problem where the message is at the center (at 0) and must be delivered to both endpoints of the segment $[-1,+1]$. 
Here we give an FPTAS in the offline case and an online algorithm that attains a competitive ratio of $\frac 95$, which we show is tight. 

\keywords{
    Broadcast \and Delivery \and Mobile Robots \and Lower Bound 
    \and Competitive Ratio \and Pony Express 
    % \and Face-To-Face \and Speed
}
\end{abstract}

\section{Introduction}
\label{sec:intro}
The Pony Express refers to the well-known mail delivery service performed by continuous horse-and-rider relays between a source and a destination point. 
It was employed in the US for a short period (1860 to 1861) to deliver mail between Missouri and California. 

The problem considered in this paper is motivated by the above.
If one thinks of the horses as  robots of differing speeds operating over a continuous domain then the Pony Express can serve as a suitable paradigm for message delivery from a source to a destination by robots passing messages from one robot to the next upon contact.
In particular, consider the following problem:
Initially a piece of information is placed at a certain location, referred to as the source. 
A group of robots are required to deliver the information from the source to another location referred to as the destination. 
The problem is one of designing a message delivery algorithm that delivers the message by selecting a sequence of robots and their movements
that relay 
the message from a source to a destination in optimal time.  

As will be seen, designing such algorithms can be a challenge given that the robots do not necessarily have the same speed and the overall delivery time may depend on what knowledge the agents possess concerning the location and speeds of the other robots. 
Further, the communication exchange model is face-to-face (F2F) in that two robots can exchange a message only when they are at the same location at the same time.   

The problem itself can be studied over any domain. In this paper, we restrict our attention to a finite interval which already offers some interesting
questions to resolve. 

\subsection{Model}
\label{sec:model}
We consider a set $R$ of $n$ robots initially scattered along a finite interval.
Each robot $r$ has a speed $v(r)$ and unique initial position $p(r)$.
Note that robots with the same initial position can be handled through some tie-breaking mechanism, adding minor perturbations to the robots' positions, etc. 
The goal is to use the robots to deliver a message to one or both of the interval endpoints.
Robots acquire the message through face-to-face contact either with the message at its initial location, or by encountering another robot with the message.
We consider three variants of the Pony Express problem in which the message is initially placed at the point $0$.

\begin{enumerate}
    \item  {\bf Pony Express.} %($\PonyExpress{1}{n}$) 
    On the interval $[0,1]$, the message must reach the endpoint $1$.
    \item {\bf Half-Broadcast.} %($\HalfBroadcast{-1}{1}{n}$) 
    On the interval $[-1,1]$, the message must reach one of the endpoints $\pm 1$.
    \item {\bf Broadcast-Problem.}  %($\Broadcast{-1}{1}{n})$  
    On the interval $[-1,1]$, the message must reach both endpoints $\pm 1$.
\end{enumerate} 

In each case, the goal is to solve the problem in the minimum amount of time. 
We consider both the offline and online settings. In the offline setting, all information regarding the robots (their initial positions and speeds) are 
available and a centralized algorithm provides a sequence of robot meetings that relay the message from the source to the destination(s) in
optimal time.

In the online setting, 
we consider a model where robots do not know their own location nor their own speeds. 
Further, the agents do no have any information about other agents (initial positions, speeds) or even the number of robots in the system.
The robots do however know the direction of the origin from their current position.
When two robots meet, they can compare their speeds and decide which is faster.

To measure the performance of our online algorithms, we consider their competitive ratios. 
Let $t^*(I)$ be the optimal delivery time for an instance $I$ of a given problem and $t_A(I)$ be the time needed by some online algorithm $A$ for the same instance.
Then the competitive ratio of $A$ is $$\max_I \frac{t_A(I)}{t^*(I)}.$$
Our goal is to find online algorithms that minimize this competitive ratio.

\subsection{Related work}
\label{sec:related}
There are many applications in a communication network where message passing (see \cite{chalopin2006mobile}) is used by agents so as to solve such 
problems as search, exploration, broadcasting and converge-casting, connectivity, and area coverage. 
For example, the authors of~\cite{beregrobustness} address the issue of how well a group of collaborating robots with limited communication range is able to monitor a given geographical space.
In particular, they study broadcasting and coverage resilience, which refers to the minimum number of robots whose removal may disconnect the network and result in uncovered areas, respectively. 
Another application may be patrolling whereby many agents are required to keep perpetually moving along a specified domain so as to minimize the time a point in the domain is left unvisited by an agent, e.g., see~\cite{czyzowicz2019patrolling} for a related survey.

%The Pony Express discussed above serves as motivation and a suitable paradigm which is consistent with the goals of message delivery by mobile agents presented in the present paper.
%There are a few related papers with related themes that have been studied recently and which we now discuss.

A general energy-aware data delivery problem was posed by~\cite{anaya2012collecting}, whereby $n$ identical, mobile agents equipped with power sources (batteries) are deployed in a weighted network. 
Agents can move along network edges as far as their batteries permit and use their batteries in linear proportion to the distance traveled. 
At the start the agents possess some initial information which they can exchange upon meeting at a node. 
The authors investigate the minimal amount of power, initially available to all agents, necessary so that convergecast may be achieved. 
They study the question in the centralized and the distributed setting. 

Two related communication problems: data delivery and convergecast are presented for a centralized scheduler which has full knowledge of the input in \cite{czyzowicz2016communication}. The authors show that if the agents are allowed to exchange energy, both problems have linear-time solutions on trees but for general undirected and directed graphs they show that these problems are NP-complete.

A restricted version of the problem above concerns $n$ mobile agents of limited energy that are placed on a straight line and which need to collectively deliver a single piece of data from a given source point $s$ to a given target point $t$ on the line.
In \cite{chalopin2013data} the authors show that the decision problem is NP-hard for a single source and also present a 2-approximation algorithm for the problem of finding the minimum energy that can be assigned to each agent so that the agents can deliver the data. 
In \cite{chalopin2014data} it is shown that deciding whether the agents can deliver the data is (weakly) NP-complete, while for instances where all input values are integers, a quasi-, pseudo-polynomial time algorithm in the distance between $s$ and $t$ is presented. 

Additional research under various conditions and topological assumptions can be found in \cite{bartschi2017truthful} which studies the game-theoretic task of selecting mobile agents to deliver multiple items on a network and optimizing or approximating the total energy consumption over all selected agents, in \cite{bartschi2017energy} which studies data delivery and combines energy and time efficiency, and in \cite{das2015collaborative,das2018collaborative} which is concerned with collaborative exploration in various topologies.

The focus of our current study is  on finding offline and online algorithms for message delivery from a source to a destination on a line segment where the goal is to minimize the time needed. This differs from the work outlined above which focuses on energy transfer and consumption to perform either a delivery or  broadcast. To the best of our knowledge, the problem and analysis considered in this paper has not been considered before.

\subsection{Outline and results of the paper}
\label{sec:outline}
In Section~\ref{sec:pony} we discuss the Pony Express variant of the problem and present optimal online and offline algorithms.
In Section~\ref{sec:half_broadcast}, we discuss the Half-Broadcast variant of the problem.
We provide an optimal offline algorithm and an online algorithm with a $\frac 32$ competitive ratio and show this ratio is the best possible.
In Section~\ref{sec:broadcast}, we discuss the Broadcast variant of the problem and provide an online algorithm with a competitive ratio of $\frac 95$
which we show is the best possible.
We also present an offline FPTAS for the Broadcast variant. (Note: the offline algorithm is not exact but depends upon performing binary search over
a real interval).
% Due to space limitations some proofs are omitted.
% A complete version of the paper can be found on arXiv.

\section{Pony Express}
\label{sec:pony}
In this section, we discuss the solution for the Pony Express variant of the problem over the segment $[0,1]$, wherein the message is placed initially at $0$ and must be delivered to $1$.

\subsection{Online}
First, we propose  an online algorithm for the Pony Express variant. The robots start at the same time and move towards the origin. The first robot to reach $0$ acquires the message.  A slower robot with the message meeting a faster one, transfers the message to the faster which then moves towards $1$.  The algorithm is as follows.  

% \begin{toappendix}
    \begin{algorithm}[H]
        \caption{Pony Express Online Algorithm}\label{alg:pony_online}
    {\small    
        \begin{algorithmic}[1]
            \State  {All robots start at the same time and move with their own speeds towards the endpoint $0$;}
            \If {a robot $r$ reaches $0$}
                \State robot $r$ acquires the message and moves towards $1$; 
            \EndIf
            \If {a robot with the message $r$ meets a robot $r^\prime$ such that $v(r) < v(r^\prime)$}
                \State {robot $r$ transmits the message to robot $r^\prime$;}
                \State {robot $r^\prime$ changes direction and moves towards $1$;}
            \Else
                \State {continue moving;}
            \EndIf
            \State {Stop when destination $1$ is reached;}
        \end{algorithmic}
    }    
    \end{algorithm}
% \end{toappendix}
 
% An Example of the Pony Express problem with the trajectories of four robots $r_0, r_1, r_2, r_3$ is depicted in Figure~\ref{fig:single_ex.pdf}.  

% % \begin{toappendix}
%     \begin{figure}[!htb]
%         \begin{center}
%             \includegraphics[width=0.8\textwidth]{FIG/single_ex.pdf}
%         \end{center}
%         \caption{Example of the Pony Express problem. Note that since $r_3$ surpasses $r_2$ before it encounters the message, $r_2$ is not necessary to deliver the message in optimal time.}\label{fig:pony_example}
%         \label{fig:single_ex.pdf}
%     \end{figure}
% % \end{toappendix}

Next we prove the optimality of Algorithm~\ref{alg:pony_online}.

\begin{theorem}
    \label{thm:pony_online}
    Algorithm~\ref{alg:pony_online} delivers the message in optimal time.
\end{theorem}
\begin{inlineproof}
    Let $m_i$ be the $i^\text{th}$ handover point between robots $r_{i-1}$ and $r_i$ at time $t_i$.
    Observe that since $r_{i-1}$ participated in the $i-1^\text{th}$ handover, it must be slower than $r_i$, or $v(r_{i-1}) < v(r_i)$ (or else there would not be a handover).
    For simplicity and consistency of notation, let $r_0$ be an additional robot with initial position and velocity $0$.
    In other words, robot $r_0$ simply holds the message at its initial position until robot $r_1$ arrives at $0$ to perform the first handover.
    Note this does not change the problem at all, since robot $r_0$ will not carry the message any distance.
    
    We show by induction and use the following inductive hypothesis: ``Each participating robot $r$ to the left of $m_i$ has speed $v(r) < v(r_i)$ and $t_i$ is the earliest time the message can be delivered to $m_i$.''
    We say the message is \textit{delivered} to a point $m$ as soon as any robot that has acquired the message (excluding the additional robot $r_0$) reaches point $m$.

    For the base case consider $m_1$. 
    Observe robot $r_1$ is the first robot to reach the source, since any robot with speed greater than $0$ would satisfy the condition to participate in a handover.
    It is clear then, that every robot to the left of $m_1 = 0$ has speed less than $v(r_1)$, since otherwise it would not be the first to arrive at $0$.
    Also, $t_1$ is the first time the message is \textit{delivered} to $0$ since $r_1$ is the first robot to arrive at $0$.

    Assume the inductive hypothesis holds for $m_{i-1}$.
    Observe that since there is a handover at $m_i$, $v(r) < v(r_{i-1}) < v(r_{i})$
    for all robots $r$ to the left of $r_{i-1}$. 
    Therefore, all participating robots to the left of $m_i$ are slower than $r_{i-1}$.
    Furthermore, the message reaches $m_i$ at the earliest possible time, since otherwise a slower robot must not have handed the message over to a faster available robot.
    Finally, observe that it is the fastest robot delivers the message to the destination point.
\qed
\end{inlineproof}

%Next, we characterize which robots for a given problem instance actually need to participate to deliver the message in optimal time.
%\begin{corollary}
%    \label{cor:pony_order}
%    A robot $r$ need only participate in delivering the message if it is faster than all robots with initial position closer to the message.
%\end{corollary}
%
%\begin{appendixproof} (Corollary~\ref{cor:pony_order}) 
%    Let $r$ and $r^\prime$ be robots such that $v(r) > v(r^\prime)$ and $p(r) < p(r^\prime)$.
%    This means $r$ is faster than $r^\prime$ and also closer to the message.
%    Assuming $r$ must participate in delivering the message, it will acquire the message before $r^\prime$.
%    Observe any robot $\hat{r}$ that $r$ hands the message over to must have speed $v(\hat{r}) > v(r) > v(r^\prime)$, and so robot $r^\prime$ will first encounter the message with a faster robot, and \textit{not} participate in a handover.
%\qed
%\end{appendixproof}

\subsection{Offline}\label{sec:pony_offline}

In this section, we present an offline algorithm for computing the optimal delivery time for the Pony Express variant (see Algorithm~\ref{alg:pony_offline}). 
In the previous section, we discussed the behavior of the robots in an optimal solution (i.e. they move toward $0$ until encountering the message and then turn around and move toward the endpoint).
The goal for an offline algorithm, then, is to compute all the meeting points where a handover occurs.
We could consider all $n^2$ possible meeting points, but that would be inefficient.
The key observation is that every robot must encounter one of its neighbors (either from its left or from its right) before encountering any other robot.
When two robots meet, either both robots are moving toward $0$ (and neither have the message) or one robot is traveling toward the endpoint with the message and the other is traveling toward $0$ to acquire it.
In either case, the meeting robots' neighbors and/or directions change, so new meeting points must be computed.
This is the idea behind the algorithm.
We keep track of potential $O(n)$ meeting points in a priority queue and examine them one-by-one to see how they affect the system.

\begin{theorem}
    \label{thm:pony_offline}
    Algorithm~\ref{alg:pony_offline} finds an optimal solution to the Pony Express problem and runs in $O(n \log n)$ time.
\end{theorem}

\begin{inlineproof}
    Observe that \texttt{q} is a Priority Queue whose operations \texttt{add(r, p)} for adding element \texttt{r} with a priority \texttt{p},
    \texttt{remove(r)} for removing element $r$ from the queue,
    \texttt{remove\_front()} for removing and returning the element with the highest priority, and \texttt{update(r, p)} for updating an element's priority in the queue each have a time-complexity of $O(\log n)$.

    In the first step of the algorithm, each robot is added to the priority queue, using its meeting time with its left-hand neighbor as a priority.
    Note this could be $\infty$ if the robot's left-hand neighbor either does not exist or moves at a faster speed away from it.
    This step has time-complexity $O(n \log n)$.

    Next, notice on each iteration of the loop in line~\ref{line:mainloop}, the size of the queue is decremented by at least one and thus terminates after 
    at most $n$ iterations.
    Therefore this part of the algorithm has time complexity $O(n \log n)$.

%    The overall complexity of the algorithm is $O(n \log n + n \log n) = O(n \log n)$.
    
    Finally, observe that robots change direction if and when they meet a slower robot with the message and meeting times are updated appropriately when a change in direction occurs. This behavior is equivalent to that of Algorithm~\ref{alg:pony_online} and therefore is optimal by Theorem~\ref{thm:pony_online}. 
\qed
\end{inlineproof} 

Algorithm~\ref{alg:pony_offline} returns only the final delivery time of the message to its destination.
Observe though, that the algorithm could easily be made to return the entire sequence of handover meeting times (each \texttt{r.meet} in line~\ref{line:handover}).

% \begin{toappendix}
    \begin{algorithm}[H]
    \caption{Pony Express Offline Algorithm}\label{alg:pony_offline}
    \hspace*{\algorithmicindent} \textbf{Input:} r, array of $n$ robot structs sorted by initial position, r[$i$].p
    {\small
    \begin{algorithmic}[1]
        \State {q $\gets$ PriorityQueue()}
        \State left $\gets$ Robot(p=$0$, v=$0$, meet=$\infty$) \Comment Additional robot to represent source
        \For {$i \gets 1 \ldots n$}
            \If {r[$i$].v $>$ left.v}
                \State r[$i$].meet $\gets \frac{\text{r[}i\text{].p }-\text{left.p}}{\text{r[}i\text{].v }-\text{left.v}}$ \Comment Meeting time when moving the same direction
            \Else
                \State r[$i$].meet $\gets \infty$
            \EndIf
            
            \If {$i \leq n - 1$}
                \State r[$i$].right $\gets$ r[$i+1$]
            \EndIf
            \State r[$i$].left $\gets$ left
            \State left $\gets$ r[$i$]
            \State q.add(r, -r[$i$].meet) \Comment Add robot to queue with meet-time-based priority
        \EndFor
    
        \State dst $\gets$ Robot(p=$1$, v=$0$, meet=$\infty$)  \Comment Additional robot to represent destination
        \State q.add(dst, $-\infty$)
        
        \While{$q.size > 0$}\label{line:mainloop}
            \State r $\gets$ q.remove\_front() \Comment Get robot with first meeting time
    
            \If {r.left.has\_message}
                \State r.has\_message $\gets$ True \label{line:handover}
                \If{r.left.v $\leq$ r.v}
                    \State q.remove(r.left)
                \EndIf
    
                \If {r.right}
                    \If{r.left.v $\leq$ r.v}
                        \State r.right.left $\gets$ r 
                    \Else
                        \State r.left.right $\gets$ r.right 
                        \State r.right.left $\gets$ r.left
                    \EndIf
                    \State r.right.meet $\gets \frac{\text{r.right.p} - \text{r.p} + 2 \cdot \text{r.meet} \cdot \text{r.v}}{\text{r.v} + \text{r.right.v}}$ \Comment Compute new meeting time
                    \State q.update(r.right, -r.right.meet)
                \EndIf

            \Else \Comment robot r passes non-participating robot
                \State q.remove(r.left)
                \State r.left $\gets$ r.left.left
                \If{r.left}
                    \State r.left.right $\gets$ r
                \EndIf
                \If{r.left.has\_message} 
                    \State r.meet $\gets \frac{\text{r.p} - \text{r.left.p} + 2 \cdot \text{r.left.meet} \cdot \text{r.left.v}}{\text{r.left.v} + \text{r.v}}$
                \ElsIf{r.v $>$ r.left.v}
                    \State r.meet $\gets \frac{\text{r.p} - \text{r.left.p}}{\text{r.v} - \text{r.left.v}}$
                \Else 
                    \State r.meet $\gets \infty$
                \EndIf
                \State q.add(r, -r.meet)
            \EndIf
        \EndWhile
        \State \textbf{return} dst.meet
    \end{algorithmic}
    }
    \end{algorithm}
% \end{toappendix}

\section{Half-Broadcast}
\label{sec:half_broadcast}
In this section we consider the Half-Broadcast variant of the problem in which a message initially placed at $0$ must be delivered to one of the endpoints of the interval $[-1, 1]$. 

\subsection{Online}
First, we show  a lower bound of $\frac 32$ on the competitive ratio for any algorithm to solve this problem.
%This lower bound also applies to models where robots have nearly full knowledge (they know the delivering robot's starting position is either $1$ or $-1$).

\begin{theorem}
    \label{thm:hb_lower}
    The competitive ratio for the Half-Broadcast problem is at least $\frac 32$.
    % i.e., $CR(\HalfBroadcast{-1}{1}{2}) \geq \frac{3}{2}$. This notation has not been introduced?
\end{theorem}
    
    \begin{inlineproof}
        Consider two robots $r$ and $r^\prime$ with speeds $v(r) = \frac 12$ and $v(r^\prime) = 1$. 
        Initially $r$ is placed at $p(r) = 0$. 
        The initial position of $r^\prime$, $p(r^\prime)$ will be determined below. 
        Let $A$ be any online algorithm for two robots. 
        Observe the movement of $r$ during the time period $[0,1]$. 
        Without loss of generality, assume that the final position of $r$ in this time period is $x \in \left[ 0, \frac{1}{2} \right]$. 
        In this case, we let $p(r^\prime) = -1$.
        (Note that if $r$ ends up in $\left[ 0 , -\frac{1}{2} \right]$, 
        we let $p(r^\prime) = 1$ and a symmetric argument will follow.)
        
        Observe that the trajectories taken by $r$ and $r^\prime$ cannot overlap during the time period $[0,1]$. 
        Indeed, at time $0 \leq t \leq 1$, $r$ is in the range 
        $\left[ x-\frac{1-t}{2}, x+\frac{1-t}{2} \right]$ (as it must reach $x$ by time 1) and $r^\prime$ is in the range $[-1, -1+t]$.
        These ranges do not overlap for $x \in \left[ 0, \frac{1}{2} \right]$ and 
        $t \in [0,1]$ 
        except at $x=0$ and $t=1$, in which case both robots are at $0$ at time $1$. 
        Thus it is not possible for $r^\prime$ to receive the message before time $1$. 
        At time 1, $r$ is at $x \in \left[ 0, \frac 1 2 \right]$ and can not make it to either endpoint ($-1$ or $1$) sooner than time $2$. 
        Let the position of $r^\prime$ be $-y \in [-1, 0]$ at time $1$. 
        If $r^\prime$ receives no help from $r$ when delivering the message then it cannot obtain the message before an additional $y$ units of time to travel from $-y$ to the message source $0$ and $1$ unit of time to bring the message to either endpoint, i.e.,  $2+y \geq 2$ ($y \geq 0$ units of time). 
        If $r^\prime$ does receive help from $r$, it cannot receive the message before time $\frac{x-y}{\frac{3}{2}}$
        and deliver the message before time 
        $1 + 1 + \frac{2(x-y)}{\frac{3}{2}} \geq 2$ for $x \in [0, \frac{1}{2}]$ and $y \in [-1,0]$. 
        
        Thus any online algorithm $A$ must take time at least $2$ units of time to solve this instance of the problem. 
        But the optimal offline algorithm can complete the task in time $\frac{4}{3}$ by having the two robots meet at time $\frac{2}{3}$ at position $-\frac{1}{3}$ and then having $r^\prime$ deliver the message to $-1$.
        Therefore the competitive ratio for any algorithm is at least $\frac{3}{2}$.
\qed
\end{inlineproof}

Next we provide an online algorithm that achieves the competitive ratio $\frac 32$. 
We consider the very simple algorithm that essentially partitions the line segment (and robots) into two instances of the Pony Express Problem (over $[-1, 0]$ and $[0, 1]$) solves them independently.
The delivery time is given by whichever instance delivers the message first.

% \begin{toappendix}
    % \vspace{-0.3cm}
    \begin{algorithm}[H]
    \caption{Half-Broadcast Online Algorithm}\label{alg:hb_online}
    {\small
        \begin{algorithmic}[1]
            \State  {All robots start at the same time and move within their own subinterval at their own speeds towards the endpoint $0$;}
            \If {a robot $r$ reaches $0$}
                \State robot $r$ acquires the message and moves towards the endpoint closest to its original position; 
            \EndIf
            \If {a robot with the message $r$ meets a robot $r^\prime$ such that $v(r) < v(r^\prime)$}
                \State {robot $r$ transmits the message to robot $r^\prime$;}
                \State {robot $r^\prime$ changes direction and moves towards the nearest endpoint;}
            \Else
                \State {continue moving;}
            \EndIf
            \State {Stop when either endpoint is reached by robot;}
        \end{algorithmic}
    }    
    \end{algorithm}
% \end{toappendix}

First, we show that Algorithm~\ref{alg:hb_online}  guarantees a competitive ratio of $\frac{3}{2}$ when only two robots participate. 
Then, we extend the result to systems of $n$ robots. Note that our algorithm is clearly optimal in the case where there is only one robot.
\begin{lemma}
    \label{lemma:hb_upper}
    Algorithm~\ref{alg:hb_online} solves the Half-Broadcast problem for the case $n=2$ with competitive ratio at most $\frac 32$.
    % i.e., $CR(\HalfBroadcast{-1}{1}{2}) \leq \frac{3}{2}$.
\end{lemma}

\begin{inlineproof} 
    Consider two robots $r$ and $r^\prime$.
    Without loss of generality, assume that $v(r) \leq v(r^\prime)$ and that in the optimal algorithm the message is delivered at $1$. 
    Considering an optimal algorithm, observe that either robot $r^\prime$ delivers the message without any help or both collaborate to deliver the message.
    In the second case, suppose that $m$ is the optimal meeting point between robots $r$ and $r^\prime$. 
    Since $v(r) \leq v(r^\prime)$, $r^\prime$ must deliver the message. 
    Thus, the delivery time is at least 

    $$
        \min \left( \frac{|p(r)| + 1}{v(r)}, \frac{|p(r^\prime)| + 1}{v(r^\prime)},  \frac{m + |p(r)|}{v(r)} + \frac{1 - m}{v(r^\prime)} \right).
    $$
    
    Observe that in cases where either robot delivers the message without collaboration, Algorithm~\ref{alg:hb_online} is optimal. 
    If the optimal algorithm requires the two robots to collaborate, however, Algorithm~\ref{alg:hb_online} is not optimal.   
    Observe that the algorithm terminates when either of the two robots arrives at an endpoint. 
    Therefore, the delivery time is maximized when
    $\frac{|p(r)| + 1}{v(r)} = \frac{|p(r^\prime)| + 1}{v(r^\prime)}$. 
    Thus, the competitive ratio is given by:
    \begin{eqnarray*}
        \frac{\frac{|p(r^\prime)| + 1}{ v(r^\prime)}  }{ \frac{m + |p(r)|}{v(r)} + \frac{1 - m}{v(r^\prime)}}  & = & \frac{|p(r^\prime)| + 1} { \frac{v(r^\prime)}{v(r)}  (m + |p(r)|) + 1 -m} \\ 
        & = & \frac{v(r)(|p(r^\prime)| + 1)} { v(r^\prime) (m + |p(r)|) + v(r)(1 -m)} \\ 
        & = & \frac{v(r)(m + |p(r^\prime)|) + v(r)(1-m)} { v(r^\prime) (m + |p(r)|) + v(r)(1 -m)}
    \end{eqnarray*}
    Observe that $p(r) = 0$ and $p(r^\prime) = 1$ maximizes the ratio.
    Thus, the competitive ratio is at most $\frac{2v(r)} { mv(r^\prime) + v(r)(1 -m)}$ and $v(r^\prime) = 2 v(r)$. Then:
    \begin{eqnarray*}
        \frac{2v(r)} { mv(r^\prime) + v(r)(1 -m)} & = & \frac{v(r^\prime)} { mv(r^\prime) + (1 -m)\frac{v(r^\prime)}{2}} \\
        & = &  \frac{2v(r^\prime)} { v(r^\prime)(m+1)} \\  
        & = & \frac{2}{m+1} \\
    \end{eqnarray*}
    Since $2v(r) = v(r^\prime)$, $m = \frac 13$ and the competitive ratio is bounded by $3/2$ for any chosen speed of $v(r^\prime)$. 
\qed
\end{inlineproof}

Now we are ready to present the main result of the section in the following theorem. We show that the competitive ratio of Algorithm~\ref{alg:hb_online} for the Half-Broadcast problem is at most $\frac 32$ when $n$ robots are participating. By Theorem \ref{thm:hb_lower}, this is best possible.

\begin{theorem}
    \label{thm:hb_online}
    The competitive ratio of Algorithm~\ref{alg:hb_online} is at most $\frac 32$ for systems of $n$ robots.
    % i.e., $CR(\HalfBroadcast{-1}{1}{n}) \leq \frac{3}{2}$.
\end{theorem}

\begin{inlineproof}
    Without loss of generality, assume that the message is delivered to $1$ in both the online and optimal offline algorithm. (Otherwise a symmetric argument can be used.) % added this part
    Let  $\mu_1, \mu_2, \ldots, \mu_k$ be the $k < n$ meeting points of the optimal centralized algorithm 
    where robot $r_{\pi(i)}$ carries the message between $\mu_{i-1}$ and $\mu_i$. Let $\mu_0 = 0$. % be the initial position of robot $r_{\pi(1)}$. 
    Let $m_1, m_2, \ldots m_l$ be the $l < n$ meeting points of the Algorithm~\ref{alg:hb_online}
    where robot $r_{\sigma(i)}$ traverses between  $m_{i-1}$ and $m_i$. Let $m_0 = 0$. % be the initial position of robot $r_{\sigma(1)}$. 
    The competitive ratio of our algorithm is

    \begin{align*}
        \frac{
            \frac{p(r_{\sigma(1)})}{v(r_{\sigma(1)})} +
            \frac{1-m_l}{v(r_{\sigma(l)})} + 
            \sum_{i=1}^{l-1}  \frac{m_i - m_{i-1}}{v(r_{\sigma(i)})}
        }{  
            \frac{p(r_{\pi(1)})}{v(r_{\pi(1)})} +
            \frac{1-\mu_k}{v(r_{\pi(k)})} + 
            \sum_{i=1}^{k-1}  \frac{\mu_i - \mu_{i-1}}{v(r_{\pi(i)})} 
        }
    \end{align*}

    % From Lemma~\ref{lemma:hb_deliver}, $r_{\sigma(l)} = r_{\pi(k)}$. 
    Observe that $m_l \leq \mu_k$ since Algorithm~\ref{alg:hb_online}
    does not attain optimal time. Therefore, $\frac{1-m_l}{v(r_{\sigma(l)})}  \leq \frac{\mu_k - m_l}{v(r_{\pi(k-1)})} +    \frac{1-\mu_k}{v(r_{\pi(k)})}$ since
    $v(r_{\pi(k-1)}) < v(r_{\pi(k)})$. 
    Observe then, that we can trim the interval at $m_l$ and solve the problem with $n-1$ robots. 
    The key observation is that the online algorithm is actually ``faster'' at each intermediate handover \textit{except} for the first handover.
    In other words, the first handover is the only segment that hurts the online algorithm.
    We have shown that the competitive ratio of the new problem is less than or equal to the competitive ratio of the original problem and
    therefore by induction (with Lemma \ref{lemma:hb_upper} as the base case) the result follows.
\qed
\end{inlineproof}

%\begin{corollary}
% There exists a local algorithm that attains optimal competitive ratio. 
%\end{corollary}
% 
%\begin{appendixproof}
% The proof follows directly from  Lemma~\ref{thm:hb_lower} and Theorem~\ref{thm:hb_online}.
%\qed
%\end{appendixproof}

\subsection{Offline}

Next, we show an offline algorithm for computing the optimal solution.
To do this, we make use of Algorithm~\ref{alg:pony_offline}, the offline algorithm for the Pony Express problem.

First we need the following lemma:

\begin{lemmarep}
    \label{lemma:hb_deliver}
    Let $r$ and $r^\prime$ be the fastest robots in the subintervals $[-1, 0]$ and $(0, 1]$, respectively.
    Then either $r$ or $r^\prime$ will deliver the message in optimal time.
\end{lemmarep}
\begin{appendixproof}
    Observe that the message must travel to one of the endpoints and therefore must traverse the entirety of one of the half-intervals.
    This means the message must encounter every robot on that half-interval, and so either $r$ or $r^\prime$ must encounter the message.

    Suppose, without loss of generality, that the message is delivered to $1$ in an optimal solution.
    Then consider the moment that $r^\prime$ encounters the message with another robot $\hat{r}$.
    Observe that, since $r^\prime$ is the fastest robot with initial position on $(0, 1]$, $\hat{r}$ can only be faster than $r^\prime$ if its initial position was on $[-1, 0)$, in which case $\hat{r}$ could deliver to~$-1$ in less time, a contradiction.
\qed
\end{appendixproof}

In an optimal solution, robots that initially start in the interval $[-1,0]$ can participate in the message delivery at the point $1$, or vice-versa. 
The following lemma  shows that even if many robots that initially start in an opposite interval can participate in delivering the message, there is always an identical solution 
where only one robot in the opposite interval participates in delivering the message. 

\begin{lemmarep}
    \label{lemma:hb_helper}
    There is an optimal solution such that at most one robot from the interval that does not contain the delivered-to endpoint participates.
\end{lemmarep}

\begin{appendixproof}
    Observe that since the message is delivered to $1$, the optimal algorithm involves all participating robots moving toward the center and then, upon encountering the message (either at the center or from a carrying robot), moving towards $1$.
    Note that robots on the subinterval $[-1, 0]$ move in one direction while robots 
    on the subinterval $(0, 1]$ travel some distance toward $0$ and then turn around.
    Observe also that the fastest robot on the interval $[0, 1]$, say $r$, must deliver the message by Lemma~\ref{lemma:hb_deliver}.
    Suppose, by contradiction, robots $r^\prime$ and $r_2$ from $[-1, 0)$ participate in the optimal algorithm.
    Let $d_1 > d_2$ ($r_2$ is closer to $0$ than $r^\prime$), then $v(r^\prime) > v_2$,
    or else $r^\prime$ can never catch up to $r_2$ and will never encounter any robots that $r_2$ has not already encountered (and thus not participate).
    Furthermore, in order to participate, $r^\prime$ must be so much faster than $r_2$ that it passes $r_2$ and encounters a faster robot on the right that has not yet encountered the message.
    In this case, however, $r_2$, may as well have not participated at all.
    Therefore, $r_2$'s participation does not improve the delivery time.
    \qed
\end{appendixproof}

\begin{theorem}
    \label{thm:hb_offline}
    There exists an offline algorithm for finding an optimal solution to the Half-Broadcast problem with time-complexity $O(n^2 \log n)$.
\end{theorem}

\begin{appendixproof}
    Without loss of generality, assume that in the optimal solution, the message is delivered to $+1$ (we can compute solutions for both the problem and its reflection).
    Therefore, according to Lemma~\ref{lemma:hb_helper}, at most \textbf{one} robot with initial position on the interval $[-1, 0)$ need participate in the message's delivery.
    Observe that if a robot $l$ with initial position $p(l) \in [-1, 0)$ helps, the delivery time would be equivalent if $l$ were reflected about $0$ and all robots with speed less than $v(l)$ removed, since, in order for $l$ to participate, it must pass $0$ and encounter either the endpoint or a faster robot at a meeting point $m$ (rendering all slower robots that reach $m$ afterwards useless).
    Given the full set of robots $R$, let $v(r)$ be the speed of robot $r$, $p(r)$ be the initial position of robot $r$, and $L = \{r \in R | p(r) < 0 \}$.
    Also, let $|r|$ be a copy of the robot $r$, 
    such that $v(|r|) = v(r)$ and $p(|r|) = |p(r)|$.
    Let $\texttt{PonyExpress}(R^\prime \subseteq R)$ be the solution for the Pony Express variant of the problem with the robots in $R^\prime$.
    Observe the solution for the Half-Broadcast problem is:
    
    \begin{align*}
        \min \left(
            \texttt{PonyExpress}(R \backslash L),
            \underset{l \in L}{\min} ~ \texttt{PonyExpress}(\{|l|\} \bigcup \{r \in R | v(r) \geq v(l)\} )
        \right)
    \end{align*}
    
    This minimization is computable with time-complexity $O(n^2 \log n)$, since Algorithm~\ref{alg:pony_offline} must be run for every robot in $L$.
\qed    
\end{appendixproof}

\section{Broadcast}
\label{sec:broadcast}
In this section we study the Broadcast variant of the problem. 
Recall that in the Broadcast problem, a message initially placed at $0$ must be delivered by robots to \textit{both} endpoints of the interval $[-1, 1]$ in minimum time. We begin with the following lemma:

\begin{lemmarep}
    \label{lemma:b_helper}
    There is an optimal solution such that at most one robot participates in the message's delivery to both endpoints.
\end{lemmarep}

\begin{appendixproof}
    Suppose for the sake of contradiction that both robots $r$ and $r^\prime$ deliver the message to participating robots in each delivery (to $-1$ and to $1$).
    Let $t_{-1}$ and $t_{-1}^\prime$ be the times that $r$ and $r^\prime$ hand the message over to robots participating in delivery to $-1$.
    Similarly, let $t_{1}$ and $t_{1}^\prime$ be the times that $r$ and $r^\prime$ hand the message over to robots participating in delivery to $1$.

    Suppose, without loss of generality, that $r$ participates in the last of these four handovers for deliver to $1$, or $t_1 > t_{-1}, t_{-1}^\prime, t_{1}^\prime$.
    Then $r$ must have passed the robot that $r^\prime$ handed the message over to at $t_{1}^\prime$ (say $\hat{r}$), otherwise it would never handover the message to a new robot, and would not participate in the delivery to $1$.
    In this case, though, robot $\hat{r}$ need not participate anymore, since a faster $r$ delivers the message to a faster robot closer to $1$.
    This is a contradiction to the statement that both robots participate in both deliveries.
    \qed
\end{appendixproof}

\subsection{Online}
\label{sec:broadcast_online}
First we show that the competitive ratio of any online algorithm is at least $\frac{9}{5}$. 
%Then, we provide an online algorithm with competitive ratio at most $\frac{9}{5}$. 

\begin{theorem}
    \label{thm:br_lower}
        The competitive ratio for any Algorithm that solves the {\em Broadcast} problem is at least $\frac 95$.
        % i.e., $CR(\Broadcast{-1}{1}{2}) \geq \frac{9}{5}$.
\end{theorem}
    
\begin{inlineproof}%
    %
    % \begin{toappendix}
        \begin{figure}[h]
            \centering
            \includegraphics[width=0.9\textwidth]{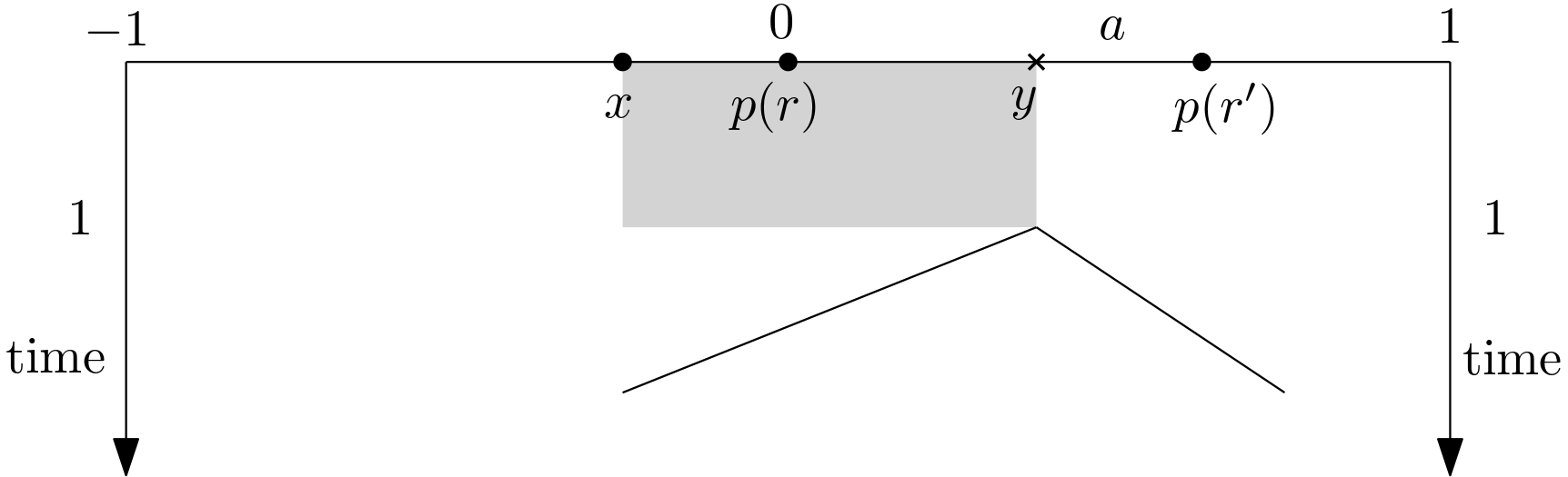}
            \label{fig:br_lower}
            \caption{
                Possible trajectory for the online algorithm. 
                Robot $r$ moves between $x$ and $y$ (the shaded region) during the time 
                interval $[0, 1]$.
            }
        \end{figure}
    % \end{toappendix}
    %
    Consider two robots, $r$ with speed $1$ and initial location $0$ and $r^\prime$ with speed and initial location to be determined below.
    Let $A$ be any online algorithm for two robots. Observe the movement of $r$ during the time period $[0,1]$ under algorithm $A$. 
    Without loss of generality, assume the final position of $r$ is $x \in [-1,0]$ and let $y$ be the furthest that $r$ progressed into $[0, 1]$ during this time period. 
    Observe that $0 \leq y \leq \frac 12$ since $r$ is in $[-1,0]$ at time 1. 
    Let $a = \frac{1-y}{2}$. 
    In this case, we set the $r^\prime$'s speed $v(r^\prime) = a$ and its initial position $p(r^\prime) = y + a$ (Figure~\ref{fig:br_lower}). 
    A symmetric argument can be used in the case that $x \in [0, 1]$.

    Observe that the trajectories of $r$ and $r^\prime$ do not overlap in the time period $[0,1]$ for any $x \in [-1,0]$ and $y \in \left[0, \frac{1}{2} \right]$ with
    the exception of the case where $y = x = 0$ and $t=1$.
    Indeed, $r^\prime$ can only reach the position $y$ at time 1. Prior to that time, its position must be to the right of $y$ and therefore to the right of $r$. 
    At time 1 it may reach $y$ but by that time $r$ is at $x \leq 0$. The only overlap
    occurs when $x = y = 0$. 

    At time 1, $r^\prime$ is at position $z \geq y \geq 0$. 
    In order to deliver to the message to either end point, it must take time 
    at least $\frac{1+z}{a} \geq \frac{2(1+y)}{1-y} \geq 2+2y$.
    Thus,  if $r^\prime$ is the first to deliver the message to one of the end points, the algorithm must take at least time $3+2y$. 
    On the other hand, if $r$ is to deliver the message to both end points, it must take at least time $y$ to reach position $y$, a further time $y$
    to return to 0, plus an additional time 3 to reach both end points. Therefore, the online algorithm $A$ must take time at least
    $3+2y$ to solve this instance of the problem. (The case where $r^\prime$ delivers the message to both endpoints is clearly worse.)

    Consider the following (offline) algorithm for the above instance: $r$ and $r^\prime$ meet at position $\frac{y+a}{1+a}$ at time $\frac{y+a}{1+a}$ (they move toward each other until meeting).
    Then $r$ delivers the message to -1 in a further $\frac{y+a}{1+a} +1$ for a total of $1+\frac{2(y+a)}{1+a}$ time. And $r^\prime$ delivers the
    message to 1 in a further $\frac{y+a}{1+a} + \frac{1-y-a}{a}= \frac{y+a}{1+a} + 1$ for a total of $1 + \frac{2(y+a)}{1+a}$ time. 

    Therefore, the competitive ratio of algorithm $A$ on this instance  is at least $\frac{3+2y}{1 + \frac{2(y+a)}{1+a}} = \frac{(3+2y)(3-y)}{5+y} \geq \frac 95$ for $y \in [0,\frac 12]$.
\qed
\end{inlineproof}

Now we show that there is an online algorithm that attains this competitive ratio.
Algorithm~\ref{alg:br_online} is very similar to the Half-Broadcast algorithm, in that we essentially partition the line segment (and robots) into two instances of the Pony Express Problem (over $[-1, 0]$ and $[0, 1]$).
The difference is that every time a robot participates in a handover (at the source, endpoint, or with another robot), it turns around and moves in the opposite direction.
This is necessary to ensure the message is delivered to \text{both} endpoints (consider the case where all robots start on one side of the message).

% \begin{toappendix}
    % \vspace{-0.3cm}
    \begin{algorithm}[H]
        \caption{Broadcast Algorithm}\label{alg:br_online}
    {\small    
        \begin{algorithmic}[1]
            \State  {All robots start at the same time and move within their own subinterval at their own speeds towards the endpoint $0$;}
            \If {a robot $r$ reaches $0$}
                \State robot $r$ acquires the message and moves towards the endpoint closest to its original position; 
            \EndIf
            \If {a robot with the message $r$ meets a robot $r^\prime$ such that $v(r) < v(r^\prime)$}
                \State {robot $r$ transmits the message to robot $r^\prime$, changes direction, and continues moving towards the opposite endpoint;}
                \State {robot $r^\prime$ changes direction and moves towards the nearest endpoint;}
            \Else
                \State {continue moving;}
            \EndIf
            \If{a robot with the message $r$ reaches the endpoint the opposite endpoint}
                \State {robot $r$ changes direction and continues moving;}
            \EndIf
            \State {Stop when both endpoints have been reached by robot;}
        \end{algorithmic}
    }   
    \end{algorithm}
% \end{toappendix}

\begin{lemmarep}
\label{thm:andtwo}
    The competitive ratio of Algorithm~\ref{alg:br_online} for the case $n=2$ is at most $\frac 95$.
    % i.e., $CR(\Broadcast{-1}{1}{2}) \leq \frac{9}{5}$.
\end{lemmarep}

\begin{appendixproof}
    Consider robots $r$ and $r^\prime$ with initial positions $p(r)$ and $p(r^\prime)$ and speeds $v(r)$ and $v(r^\prime)$, respectively.
    Without loss of generality, assume that $v(r) \leq v(r^\prime)$. 
    Observe that, in any optimal algorithm, either one robot delivers the message to both sides without collaboration, each robot delivers the message without any collaboration, or each robot delivers the message and collaborates.
    In the third case, suppose that $m$ is the meeting point. 
    Without loss of generality, we assume that $m \geq 0$. 
    Further, since $v(r^\prime) \geq 0$,  we can assume that $r^\prime$ does not stop until it meets $r$. 
    Thus, the deliver time is at most  

    \begin{align*}
        \min\left(
            \frac{|p(r^\prime)| + 3}{v(r^\prime)}, 
            \max\left( \frac{|p(r)| + 1}{v(r)}, \frac{|p(r^\prime)| + 1}{v(r^\prime)} \right), 
            \frac{m + |p(r^\prime)|}{v(r^\prime)} + \max\left(\frac{1 + m}{v(r^\prime)}, \frac{1 - m}{v(r)}\right)
        \right).
    \end{align*}

    % \begin{toappendix}
        \begin{figure}
            \centering
            \includegraphics[scale=1]{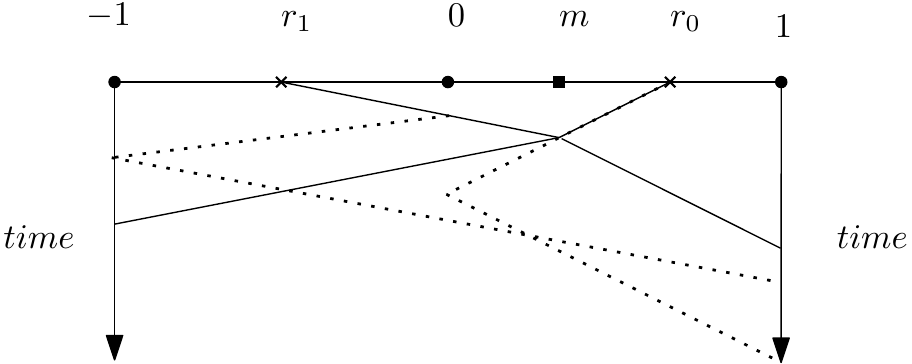}
            \caption{The dotted line represents the trajectory of the robots running Algorithm~\ref{alg:br_online} and the solid lines
            the represent the trajectory of the optimal algorithm. }
            \label{fig:andoptimal}
        \end{figure}
    % \end{toappendix}
    
    Now consider the delivery time of Algorithm~\ref{alg:br_online}. 
    Without loss of generality, assume that $p(r) > p(r^\prime)$.
    We consider two cases, the case where  $p(r) \geq 0$ and $p(r^\prime) \leq 0$ and the case where  $p(r) \leq 0$ and $p(r^\prime) \leq 0$.

    Let us first consider the case where $p(r) \geq 0$ and $p(r^\prime) \leq 0$. 
    The case where $p(r) \leq 0$ and $p(r^\prime) \geq 0$ is analogous. 
    Observe that if in the optimal algorithm both deliver the message without help or only one delivers the message to both sides without help 
    then Algorithm~\ref{alg:br_online} attains optimal deliver time. 
    It is sufficient, then, to consider the case where, optimally, each delivers the message to an endpoint \textit{after} collaboration (Figure~\ref{fig:andoptimal}). Therefore, the competitive ratio is given by

    \begin{align*}
        \frac{
            \min\left(
                \max\left(\frac{1+|p(r^\prime)|}{v(r^\prime)}, 
                \frac{1 + p(r)}{v(r)}\right), \frac{3+|p(r^\prime)|}{v(r^\prime)}
            \right)
        }{
            \frac{m + |p(r^\prime)|}{v(r^\prime)} + \max\left(\frac{1 + m}{v(r^\prime)}, \frac{1 - m}{v(r)}\right)
        }.
    \end{align*}

    Observe that the competitive ratio is maximum when 
    $\frac{m + |p(r^\prime)|}{v(r^\prime)} + \max \left(\frac{1 + m}{v(r^\prime)}, \frac{1 - m}{v(r)} \right)$
    is maximum, which happens when $\frac{1 + m}{v(r^\prime)} =  \frac{1 - m}{v(r)}$.
    Let $x$ be the equilibrium point, i.e., $\frac{1-x}{v(r)} = \frac{x + 1}{v(r^\prime)}$ and let $y = m - x$.
    Observe that $r$ initially is in the interval $[m-w, m+w] = [x+y-w, x+y+w]$ where  $w = v(r)\frac{m + |p(r^\prime)|}{v(r^\prime)}$.
    We can rewrite the competitive ratio as: 

    \begin{align*}
        \frac{
            \min\left(
                \max\left(
                    \frac{1+|p(r^\prime)|}{v(r^\prime)}, \frac{1+p(r)}{v(r)}
                \right), 
                \frac{3+|p(r^\prime)|}{v(r^\prime)}
            \right)
        }{
            \frac{x + |p(r^\prime)|}{v(r^\prime)} + \frac{y}{v(r^\prime)} + 
            \max\left(
                \frac{1 - (x+y)}{v(r)}, \frac{1 + (x+y)}{v(r^\prime)} 
            \right)
        } &= \frac{
            \min\left( 
                \frac{1 + x}{v(r)} + \frac{x+y + |p(r^\prime)|}{v(r^\prime)},
                \frac{3+|p(r^\prime)|}{v(r^\prime)}
            \right)
        }{
            \frac{x + |p(r^\prime)| + y}{v(r^\prime)}  + 
            \max\left(
                \frac{1 - (x+y)}{v(r)}, \frac{1 + (x+y)}{v(r^\prime)} 
            \right)
        } \\
        &= \frac{
            \min\left( 
                \frac{1 + x}{v(r)} + \frac{x}{v(r^\prime)} , \frac{3-y}{v(r^\prime)}
            \right) + \frac{y + |p(r^\prime)|}{v(r^\prime) }
        }{
            \frac{x}{v(r^\prime)} +
            \max\left(
                \frac{1 - (x+y)}{v(r)}, \frac{1 + (x+y)}{v(r^\prime)} 
            \right) + \frac{y + |p(r^\prime)|}{v(r^\prime)}
        }.
    \end{align*}

    Observe that the values of $y$ and $p(r^\prime)$ that maximize the competitive ratio are $y = 0$ and $p(r^\prime) = 0$ so that
    the competitive ratio is bounded by:

    \begin{align*}
        \frac{
            \min\left(
                \frac{1 + x}{v(r)} + \frac{x}{v(r^\prime)},
                \frac{3}{v(r^\prime)}
            \right)  
        }{
            \frac{x}{v(r^\prime)} + \max\left( 
                \frac{1 - x}{v(r)},
                \frac{1 + x}{v(r^\prime)} 
            \right)  
        }.
    \end{align*}

    To maximize the competitive ratio, we maximize 
    $\min \left( \frac{1 + x}{v(r)} + \frac{x}{v(r^\prime)} , \frac{3}{v(r^\prime)} \right)$ 
    by making  $\frac{1+x}{v(r)} + \frac{x}{v(r^\prime)} = \frac{3}{v(r^\prime)}$. 
    Therefore, $\frac{v(r^\prime)}{v(r)} = \frac{3-x}{1+x}$. 
    Since  $\frac{v(r^\prime)}{v(r)} = \frac{x + 1}{x-1}$, we obtain $\frac{x + 1}{1-x}  = \frac{3 - x}{1+x}$ which results in  $x = \frac 13$. 
    Hence, $\frac{v(r^\prime)}{v(r)} = 2$ and we can compute the maximum competitive ratio as

    \begin{align*}
        \frac{ 
            \frac{x}{v(r^\prime)} 
        }{
            \frac{x}{v(r^\prime)} +
            \max\left(
                \frac{1 - x}{v(r)}, 
                \frac{1 + x}{v(r^\prime)} 
            \right)  
        } = \frac{3}{
            \frac 13 + \max\left(
                \frac{\frac 23 v(r^\prime)}{v(r)}, \frac 43 
            \right)  
        } = \frac 95 .
    \end{align*}

    % \begin{toappendix}
        \begin{figure}
            \centering
            \includegraphics[scale=1]{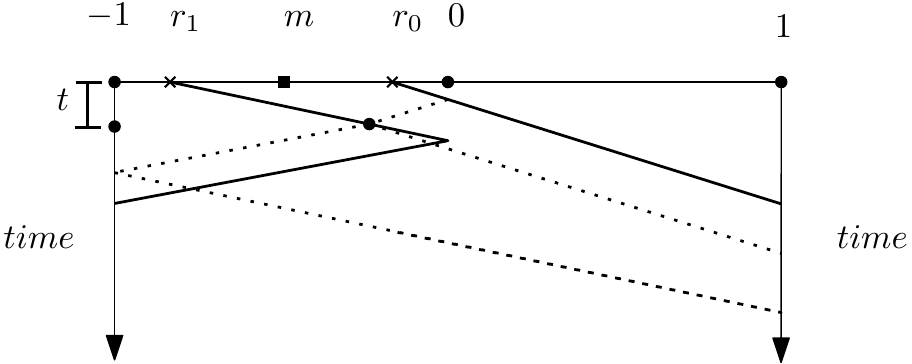}
            \caption{The dotted line represents the trajectory of the robots running Algorithm~\ref{alg:br_online} and the solid lines
            the represent the trajectory of the optimal algorithm. }
            \label{fig:br_optimal}
        \end{figure}
    % \end{toappendix}

    It remains to show the case where $p(r) \leq 0$ and $p(r^\prime) \leq 0$. 
    Observe that if in the optimal algorithm robots collaborate to deliver the message, 
    then Algorithm~\ref{alg:br_online} attains the optimal time since both are in the same segment. 
    However, if in the optimal algorithm each robot delivers the message to one side, then  Algorithm~\ref{alg:br_online} cannot attain optimal time (Figure~\ref{fig:br_optimal}).
    Therefore, the competitive ratio is given by 

    \begin{align*}
        \frac{ 
            \min\left(
                \frac{1 - p(r)}{v(r)}, 
                \frac{3 + p(r^\prime)}{v(r^\prime)}
            \right) + 2t
        }{ 
            \max\left(
                \frac{1 - p(r)}{v(r)}, 
                \frac{1 - p(r^\prime)}{v(r^\prime)} 
            \right) 
        },
    \end{align*}
    where $t$ is the time that robots need to meet and return to the original location.
    To maximize the competitive ratio, we maximize $t = \frac{|p(r) - p(r^\prime)|}{v(r) + v(r^\prime)}$ by assigning the extreme values to $r^\prime$ and $r$, i.e., 
    $r^\prime = -1$ and $r = 0$. 
    Moreover, we maximize the competitive ratio by setting 
    $\frac{1 - p(r)}{v(r)} = \frac{3 + p(r^\prime)}{v(r^\prime)}$. 
    Therefore, $\frac{v(r^\prime)}{v(r)} = \frac{3 + p(r^\prime)}{r + 1} = 2$
    and $t = \frac{1}{3 v(r)}$ and the competitive ratio is bounded by 
    
    \begin{align*}
        \frac{ 
            \frac{1 - p(r)}{v(r)} + \frac{2}{3v(r)}
        }{ 
            \max\left(
                \frac{1 - p(r)}{v(r)}, 
                \frac{1 - p(r^\prime)}{v(r^\prime)} 
            \right) 
        } = \frac{
            1 + \frac 23 
        }{ 
            \max\left( 1, 2\frac{v(r)}{v(r^\prime)} \right) 
        } = \frac 53 .
    \end{align*}
\qed
\end{appendixproof}

Next, we show that Algorithm~\ref{alg:br_online} attains optimal competitive ratio with $n\geq 3$ robots. 

\begin{theorem}
    \label{thm:br_online}
    The competitive ratio of Algorithm~\ref{alg:br_online} for systems of $n$ robots is at most $\frac 95$.
    % i.e., $CR(\Broadcast{-1}{1}{n}) \leq \frac{9}{5}$.
\end{theorem}

\begin{appendixproof}
    Let  $\mu^+_1, \mu^+_2, \ldots \mu^+_{k^+}$ and  $\mu^-_1, \mu^-_2, \ldots \mu^-_k$ be the $k^+ \leq n$  and 
    $k^- \leq n$ meeting points of the optimal centralized algorithm 
    of the robots that carry the message to $1$ and $-1$ respectively where 
    robot $r^+_{\pi(i)}$ and $r^-_{\pi(i)}$ carries the message between  $\mu^+_{i-1}$ and $\mu^+_i$ and
    $\mu^+_{i}$ and $\mu^+_{i-1}$, respectively. 
    Let $\mu^+_0$ and
    $\mu^-_0 = 0$. % be the initial position of robot $r^+_{\pi(1)}$ and $r^-_{\pi(1)}$. 

    Let  $m^+_1, m^+_2, \ldots m^+_{k^+}$ and  $m^-_1, m^-_2, \ldots m^-_k$ be the $l^+ \leq n$  and 
    $l^- \leq n$ meeting points of the Algorithm~\ref{alg:br_online}
    of the robots that carry the message to $1$ and $-1$ respectively where 
    robot $r^+_{\sigma(i)}$ and $r^-_{\sigma(i)}$  carry the message between  $m^+_{i-1}$ and $m^+_i$ 
    and $m^-_{i}$ and $m^-_{i-1}$, respectively. Let $m^+_0$ and
    $m^-_0 = 0$ % be the initial position of robot $r^+_{\sigma(1)}$ and $r^-_{\sigma(1)}$. 
    
    Then the competitive ratio of our algorithm is

    $$
    = \frac{ \max\left(
    \sum_{i=1}^{l^--1}  \frac{m^-_{i-1} - m_i}{v(r^-_{\sigma(i)})} + \frac{m^-_l + 1}{v(r^-_{\sigma(l)})},   
    \sum_{i=1}^{l^+-1}  \frac{m^+_i - m^+_{i-1}}{v(r^+_{\sigma(i)})} + \frac{1-m^+_l}{v(r^+_{\sigma(l)})}
    \right)}{ \max\left(\sum_{i=1}^{k^--1}  \frac{\mu^-_{i-1} - \mu^-_i}{v(r^-_{\pi(i)})} + \frac{\mu^-_k+1}{v(r^-_{\pi(k)})},
                \sum_{i=1}^{k^+-1}  \frac{\mu^+_i - \mu^+_{i-1}}{v(r^+_{\pi(i)})} + \frac{1-\mu^+_k}{v(r^+_{\pi(k)})}  
    \right) }. 
    $$ 
		
		We claim that   $v(r^+_{\sigma(l^+)}) \geq v(r^+_{\pi(k^+)})$ and $v(r^-_{\sigma(l^-)}) \geq v(r^-_{\pi(k^-)})$. 
Indeed, if the fastest robot in each side delivers the message in Algorithm~\ref{alg:br_online}, then 
$v(r^+_{\sigma(l^+)}) = v(r^+_{\pi(k^+)})$ and $v(r^-_{\sigma(l^-)}) = v(r^-_{\pi(k^-)})$. Otherwise, we claim that in 
Algorithm~\ref{alg:br_online}, the fastest robot delivers the message to both sides. The key observation is that when the
fastest robot gets the message, it never hands it over to another robot and that it is nearest to the endpoint that is delivered to first.
To see the latter, if the fastest robot is on the other side, it would take at most $\frac 2v$ where $v$ 
is the speed of the fastest robot. Therefore, it arrives earlier and $v(r^+_{\sigma(l^+)}) = v(r^+_{\pi(k^+)})$ and $v(r^-_{\sigma(l^-)}) = v(r^-_{\pi(k^-)})$.
Therefore, the fastest robot delivers the message in one side and then reaches the other side by itself.
%    From Lemma~\ref{lemma:hb_deliver},
 %    $r^+_{\sigma(l^+)} = r^+_{\pi(k^+)}$ and  $r^-_{\sigma(l^-)} = r^-_{\pi(k^-)}$. 
Observe that $m^+_{l^+} \leq \mu^+_{k^+}$ and $m^-_{l^-} \geq \mu^-_{k^-}$
    since Algorithm~\ref{alg:hb_online}
    does not attain optimal time. Therefore, $\frac{1-m^+_l}{v(r^+_{\pi(k)})}  \leq    \frac{1-\mu^+_k}{v(r^+_{\pi(k)})}$ and  $\frac{m^-_l-1}{v(r^-_{\pi(k)})}  \leq    \frac{\mu^-_k -1}{v(r^-_{\pi(k)})}$ since
    $v(r^+_{\pi(k^+-1))} < v(r^+_{\pi(k^+)})$ and $v(r^-_{\pi(k^--1)}) < v(r^-_{\pi(k^+-)})$. Therefore, we can trim the line at $\max \left(m^+_{l^+}, |m^-_{l^-}| \right)$ and solve the problem with $n-1$ robots and we have shown the competitive ratio of the smaller problem is less than or equal to the original problem.
    
 %   $$CR(\Broadcast{-1}{1}{n}) \leq CR(\Broadcast{-\max(m^+_{l^+}, |m^-_{l^-}|)}{\max(m^+_{l^+}, |m^-_{l^-}|)}{n-1})$$
    
    The theorem follows inductively with Lemma \ref{thm:andtwo} acting as the base case. 
        %since $CR(\Broadcast{-1}{1}{2})$ and from Theorem~\ref{thm:br_online}, the competitive ratio is bounded by $\frac{9}{5}$.
    %This completes the proof.
\qed
\end{appendixproof}

% \begin{corollary}
% \label{cor:br_online}
%     There exists a local algorithm that attains optimal competitive ratio. 
% \end{corollary}
 
% \begin{appendixproof}
%     The proof follows directly from  Lemma~\ref{thm:br_lower} and Theorem~\ref{thm:br_online}.
% \qed
% \end{appendixproof}

\subsection{Offline}
\label{sec:br_offline}
In this section, we provide an offline fully polynomial time approximation scheme (FPTAS).

\begin{theorem}
    \label{thm:br_offline}
    For any $\epsilon >0$,
    there exists an algorithm for finding a solution to within an additive factor of $\epsilon$ of optimal to the Broadcast problem with running time $O(n^2 \log n \log \frac 1\epsilon)$.
\end{theorem}

\begin{inlineproof}
    According to Lemma~\ref{lemma:b_helper}, at most \textbf{one} robot must cross $0$ and participate in the message's delivery to both endpoints.
    This robot, say $r$, may participate by delivering the message itself or handing it over to another robot.
    It's important to note that the receiving robot may not be the first encountered by $r$ nor must it be 
    faster than $r$. 
    We must consider the scenarios where $r$ delivers the message to each of the possible robots on the opposite 
    subinterval.
    To facilitate the formulation of the solution, we assume there are robots with speed $0$ at both endpoints 
    $-1$ and $1$ so that delivering to these robots is equivalent (in time and meaning) to delivering to the destination.

    Suppose the optimal handover on the opposite side of the interval occurs at position $m$ on the segment.
    Then, observe that since all robots must only participate in delivering the message to the nearest endpoint,
    there are essentially two instances of the regular PonyExpress problem to solve (one for each endpoint).
    One instance is on the interval $[-1, \min(m, 0)]$ and the other $[max(m, 0), 1]$.
    Also, the robots have shifted some distance toward $0$, based on their speeds.
    This new instance can be constructed in linear time and solved in $O (n \log n)$ time by 
    Algorithm~\ref{alg:pony_offline}.

    All that remains, then, is to find $m$. 
    Observe that two robots $l$ and $r$ with initial positions $p(l)$ and $p(r)$, and speeds $v(l)$ and $v(r)$, respectively can meet at any point on the interval
    $$\left[0, \min\left(1, \frac{p(r) - p(l)}{v(l) - v(r)}\right)\right].$$
    So the optimal solution can be described as:

    \begin{align*}
        \underset{l \in L, r \in \overline{L}}{min} ~ 
        \underset{m \in \left[
            0, \min\left(
                1, \frac{p(r) - p(l)}{v(l) - v(r)}
            \right)
        \right]}{min} ~
        \texttt{PonyExpress}\left(T\left(R, \frac{m - p(l)}{v(l)}, m\right)\right).
    \end{align*}

    Notice the inner minimization is over a real domain where the function is bitonic and so it can be estimated using binary search.
    The runtime for this is therefore  $O(n^2 \log n \log \frac 1\epsilon)$ where the computed meeting point is within $\epsilon$ of the optimal meeting point.
\qed
\end{inlineproof}

\section{Conclusion}
\label{sec:conclusion}
In this paper, we have introduced the Pony Express problem. We considered the case where the domain of interest is a line segment and the cases
where a message must be delivered from one end to the other (Pony Express), from the center to one of the end points (Half-Broadcast) and
from the center to both end points (Broadcast). For the first two problems we provide polynomial time offline algorithms and for the third an FPTAS. 
We provide online algorithms for each problem with best possible competitive ratio in each case. 

A number of open problems are suggested by our study. First, it seems likely the the runtime of our offline algorithms may be improved at least
for the case of Half-Broadcast and Broadcast and that an exact algorithm exists for Broadcast. Second, it might be worth considering variations on
the amount and type of information available to the agents in the online setting.  Finally,
another direction of study would be to consider domains other than a finite interval. 
Preliminary results for the plane can be found in~\cite{pony_plane}. 

\bibliographystyle{splncs04}
\bibliography{refs}

\end{document}